\documentclass[iop, twocolappendix, numberedappendix]{emulateapj}
   
\usepackage{amsmath}
\usepackage{aas_macros}
\usepackage{natbib}
\usepackage{xspace}
\usepackage[varg]{txfonts}
\usepackage{graphicx}
\usepackage{nicefrac}
\usepackage{multirow}
\usepackage{soul}
\usepackage{float}
\usepackage[T1]{fontenc}
\usepackage{lmodern}
 
\usepackage[dvipsnames]{xcolor}

\colorlet{mylinkcolor}{violet}
\colorlet{myurlcolor}{YellowOrange}
\colorlet{mycitecolor}{Aquamarine}

\newcommand{\Msun}{\ensuremath{\,M_\odot}\xspace}

\newcommand{\heii}{He\,{\sc ii} $\lambda$4686}

\begin{document}

\title{Ultraviolet Detection of the Binary Companion to the Type 
II\lowercase{b} SN~2001\lowercase{ig}}

\author{Stuart~D.~Ryder\altaffilmark{1}}
\altaffiltext{1}{Australian Astronomical Observatory, 105 Delhi Rd, 
North Ryde, NSW 2113, Australia.}
\email{sdr@aao.gov.au}
\author{Schuyler~D.~Van Dyk\altaffilmark{2}}
\altaffiltext{2}{Caltech/IPAC, Mailcode 100-22, Pasadena, CA 91125,
USA.}
\author{Ori~D.~Fox\altaffilmark{3}}
\altaffiltext{3}{Space Telescope Science Institute, 3700 San Martin Drive,
Baltimore, MD 21218, USA.}
\author{Emmanouil~Zapartas\altaffilmark{4}}
\altaffiltext{4}{Anton Pannekoek Institute for Astronomy, 
University of Amsterdam, Science Park 904, 1098 XH, Amsterdam,
The Netherlands.}
\author{Selma~E.~de Mink\altaffilmark{4}}
\author{Nathan~Smith\altaffilmark{5}}
\altaffiltext{5}{Steward Observatory, University of Arizona, Tucson,
AZ 85721, USA.}
\author{Emily~Brunsden\altaffilmark{6}}
\altaffiltext{6}{Department of Physics, University of York, 
Heslington, York YO10 5DD, United Kingdom.}
\author{K.~Azalee~Bostroem\altaffilmark{7}}
\altaffiltext{7}{Department of Physics, University of California, Davis,
CA 95616, USA.}
\author{Alexei~V.~Filippenko\altaffilmark{8,9}}
\altaffiltext{8}{Department of Astronomy, University of California, 
Berkeley, CA 94720-3411, USA}
\altaffiltext{9}{Miller Senior Fellow, Miller Institute for Basic 
Research in Science, University of California, Berkeley, CA 94720, USA}
\author{Isaac Shivvers\altaffilmark{8}}
\author{WeiKang Zheng\altaffilmark{8}}

\begin{abstract} 
We present {\sl HST}/WFC3 ultraviolet imaging in the F275W and F336W
bands of the Type~IIb SN~2001ig at an age of more than 14~years. A clear
point source is detected at the site of the explosion having 
$m_{\rm F275W}=25.39 \pm 0.10$ and $m_{\rm F336W}=25.88 \pm 0.13$ mag.
Despite weak constraints on both the distance to the host galaxy
NGC~7424 and the line-of-sight reddening to the supernova, this 
source matches the characteristics of an early B-type main
sequence star having $19,000 < T_{\rm eff} < 22,000$~K and 
$\log (L_{\rm bol}/L_{\odot})=3.92 \pm 0.14$. A BPASS v2.1
binary evolution model, with primary and secondary masses of 13~\Msun\
and 9~\Msun\ respectively, is found to resemble simultaneously in the
Hertzsprung-Russell diagram both the observed location of this
surviving companion, and the primary star evolutionary endpoints for
other Type~IIb supernovae.
This same model exhibits highly variable late-stage mass loss, as
expected from the behavior of the radio light curves. A Gemini/GMOS
optical spectrum at an age of 6 years reveals a narrow \heii\ 
emission line, indicative of continuing interaction with a dense
circumstellar medium at large radii from the progenitor. We review our 
findings on SN~2001ig in the context of binary evolution channels for
stripped-envelope supernovae. Owing to the uncrowded nature of its
environment in the ultraviolet, this study of SN~2001ig represents one
of the cleanest detections to date of a surviving binary companion to a
Type~IIb supernova.
\end{abstract}


\section{Introduction}\label{intro}

Core-collapse supernovae (CCSNe) can result when massive stars exhaust 
their available fuel and the cores collapse under their own weight, 
thereby releasing enough potential energy to eject their outer layers
\citep{Bethe+1979,Woosley+2002}. Stripped-envelope SNe (SESNe) are a
subset of CCSNe with progenitor stars that have lost most
or all of their outer hydrogen envelope, and in some cases even their
helium envelopes \citep[e.g.,][]{Fileppenko1997}.

Type~IIb supernovae (SNe) are distinguished by their initially strong 
H lines that fade away over the course of weeks to months 
\citep{Filippenko1988,Fileppenko1997,Filippenko1993,Gal-Yam2016}. 
This moderate amount of observed H is usually interpreted as an
intermediate case between the H-rich 
SNe IIP/IIL and H-poor SNe~Ib/c, presumably reflecting an increase in
stripping of the stellar envelope from IIP/L 
$\rightarrow$ IIb $\rightarrow$ Ib $\rightarrow$ Ic.   
Recent evidence, however, appears to demonstrate that SNe~IIb 
are spectroscopically distinct from SNe~Ib/c at all epochs
\citep{Liu2016}. Further indications that SNe~IIb form a separate
channel from SNe IIP/IIL come from their low ejecta masses
\citep{Lyman2016}, together with radiative transfer models of their spectra
\citep{Dessart2012}.

At least two scenarios can account for the stripping of the progenitor
star's envelope prior to its eventual demise in a SN explosion (for both
SNe~IIb and Ib/c subclasses). Stellar winds accompanied by extreme or
eruptive mass loss can shed significant amounts of gas over the lifetime 
of massive stars, \citep[e.g.,][]{Heger2003,Smith2006}; however the 
mass loss rates for these tend to be overestimated \citep{Smith2014b}.
Alternatively (or additionally), interaction with a massive binary
companion can transport the H-rich outer layers into the circumstellar
medium 
\citep[CSM; e.g.,][]{Podsiadlowski1993,Eldridge2008,Claeys2011,Hirai2014}.  
These two scenarios allow for a large range of potential progenitor 
systems and thus of possible companions at the moment of explosion, as 
summarized by \citet{Zapartas2017a}. In scenarios that include a surviving 
binary companion, most models predict that this survivor should be a 
relatively unevolved, hot main sequence star \citep{Eldridge2015}.

Identification of the progenitor system can provide important constraints
on the mass loss scenarios and theoretical evolution of massive stars.
While SNe~IIb constitute only $\sim$10\% of local CCSNe
\citep{Smith2011,Shivvers2017}, they are one of the best-studied
subclasses, including several progenitor detections.  SN~1993J in M81
attracted considerable interest in part owing to its proximity (3.6~Mpc),
which enabled the detection of not just the progenitor star
\citep{Aldering1994,VanDyk2002} but also a putative early B-type
supergiant companion star \citep{Maund2004,Fox2014}.  Progenitor stars
have also been identified in pre-explosion images of the Type~IIb 
SNe 2008ax \citep{Crockett2008,Folatelli2015}, 2011dh
\citep{Maund2011,VanDyk2011,VanDyk2013}, 2013df \citep{VanDyk2014}, 
and 2016gkg \citep{Tartaglia2017,Kilpatrick2017}, while searches for
a binary companion to SN~2011dh are inconclusive
\citep{Folatelli2014,Maund2015}. The relatively low initial masses
inferred from these SNe~IIb progenitors, which overlap with the
progenitor mass range of SNe~IIP \citep{Smartt2009}, further
complicate the hypothesis 
that SNe~IIb form a transition from normal SNe~IIP/IIL to SNe~Ibc 
with increasing initial mass and/or mass loss rate.

The mass loss and progenitor properties can also be constrained 
indirectly. The interaction of the expanding SN blast wave with a 
pre-existing CSM consisting of material shed by the stellar progenitor
gives rise to multi-wavelength emission, including X-ray inverse Compton
scattering and radio continuum synchrotron radiation. Multi-frequency
radio light curves on timescales of hours to years enable the 
reconstruction of the progenitor star's mass loss history centuries into
the past.  With a measured or assumed wind speed, the mass loss rate can
be calculated as a function of time to narrow down potential progenitors
\citep{Weiler2002,Horesh2013,Smith2014b}.

The Type~IIb SN~2001ig was discovered visually by \citet{Evans2001} on
2001~December~10.43~UT in the outskirts of the nearby late-type spiral
galaxy NGC~7424, but received surprisingly little optical study despite
reaching 12$^{\rm th}$ magnitude \citep{Bembrick2002}.
\citet{Silverman2009} published 12 optical spectra, 
\citet{Maund2007} presented 3~epochs of optical spectropolarimetry, and 
\citet{Ryder2006} showed a late-time spectrum, all taken within the 
first year
after explosion. \citet{Silverman2009}, \citet{Maund2007}, and
\citet{Shivvers2013} pointed out both the observed similarities and
differences between SN~2001ig and SN~1993J. \citet{Shivvers2013} further
compared SN~2001ig in the nebular phase with SN~2011dh.
Had there been an optical light curve for SN~2001ig, particularly at 
early times, it would have been evident from the timing of the decline
from the initial cooling phase whether SN~2001ig was more like SN~1993J,
with its extended-envelope progenitor, or SN~2011dh, arising from a
somewhat more compact star.

SN~2001ig was detected in X-rays with {\sl Chandra\/}
\citep{Schlegel2002}, yet the most complete coverage of all was at 
radio wavelengths. Following the initial detection within 5~days of
discovery, \citet{Ryder2004} presented almost 2~years worth of data 
from the Australia Telescope Compact Array (ATCA) and the Very Large 
Array (VLA) at frequencies between 1.4 and 22.5~GHz. Fitting the 
multi-frequency light curves enabled them to infer a mass loss rate 
from the progenitor of $\sim$$2\times10^{-5}$~M$_{\odot}$~yr$^{-1}$, 
for a wind velocity of 10~km~s$^{-1}$.

There are no pre-explosion images of NGC~7424 of sufficient depth
or resolution to enable a direct identification of the progenitor
to SN~2001ig. The radio light curves, however, exhibited regular
modulations with a period of $\sim$150~days between peaks, which
\citet{Ryder2004} attributed to ``sculpting'' of the CSM by a binary
companion. \citet{Ryder2006} imaged the location of SN~2001ig with the
Gemini Multi-Object Spectrograph (GMOS) on the Gemini South 8~m 
telescope and identified an optical counterpart in $g^{\prime}$ and
$r^{\prime}$ data, consistent with the presence of a surviving 
supergiant companion of spectral type late-B through late-F, depending 
on the extinction assumed.

\citet{Soderberg2006} drew attention to several similarities between 
the radio light curves of SN~2001ig and those of the broad-lined 
Type Ic SN~2003bg, including the strength of the modulations and 
their period. Since the likelihood of these two distinct SN subtypes
having almost the same binary progenitor system properties (orbital
period, mass ratio) that would give rise to such similar CSM structures
must be extremely small, \citet{Soderberg2006} favored instead a
scenario in which both SN~2001ig and SN~2003bg have single Wolf-Rayet
star progenitors that undergo quasi-periodic episodes of enhanced 
mass loss. \citet{Kotak2006} suggested that S~Doradus-type variations, 
as seen in Luminous Blue Variables (LBVs), could be one such mechanism.
\citet{Ben-Ami2015} analyzed early-time ultraviolet (UV) spectra
obtained with the {\sl Hubble Space Telescope\/} ({\sl HST}) of four 
SNe~IIb, including SN~2001ig. In contrast to SNe 1993J, 2011dh, and
2013df, the UV spectra of SN~2001ig showed a quite weak continuum, 
and strong reverse-fluorescence features more akin to those of
Type Ia SNe, consistent with a high radioactive $^{56}$Ni mass and 
a compact progenitor object.

As part of {\sl HST\/} program GO-14075 
(PI: O.~Fox), we used the Wide Field Camera 3 (WFC3) to search 
for surviving binary companions to nearby SESNe in the UV. 
In \citet{Zapartas2017b} we placed deep upper limits on the mass of a
binary companion to the broad-lined Type Ic SN~2002ap, and used binary
population synthesis models to rule out the 40\% of SESN channels that
would have resulted in a surviving main sequence companion more massive
than this. Section~\ref{sec:hst} presents our new {\sl HST}/WFC3 UV
observations of SN~2001ig, while Section~\ref{sec:gmos} describes
previously unpublished Gemini South/GMOS spectroscopy from 2007.  
We outline our photometric results in Section~\ref{sec:photres} and 
our spectroscopic results in Section~\ref{sec:specres}. Our
interpretation of these in terms of a binary progenitor system is in
Section~\ref{sec:disc} and our conclusions are summarized in
Section~\ref{sec:con}. We note that NGC~7424 has also recently served 
as host to the Type~II SN~2017bzb \citep{Morrell2017}.

\section{Observations}
\label{sec:obs}

\subsection{{\sl HST\/} Imaging}
\label{sec:hst}

We imaged the site of SN~2001ig with WFC3/UVIS on 2016 April 28 UT
(14.4 yr after explosion) in bands F275W and F336W, with total 
exposures of 8694~s and 2920~s, respectively. The exposures were 
line-dithered, to improve image quality and resolution and to mitigate
against hot pixels and cosmic ray hits, and post-flashed to help
mitigate against charge-transfer efficiency (CTE) losses. The images 
had been processed through the standard pipeline at the Space 
Telescope Science Institute (STScI) before we obtained them from the
Mikulski Archive for Space Telescopes (MAST). Specifically, the
individual {\em flt} frames had been corrected for CTE losses. 
We then ran these corrected {\em flc} frames through the routine
AstroDrizzle within PYRAF, to create final image mosaics in each band.

\begin{figure}
\plottwo{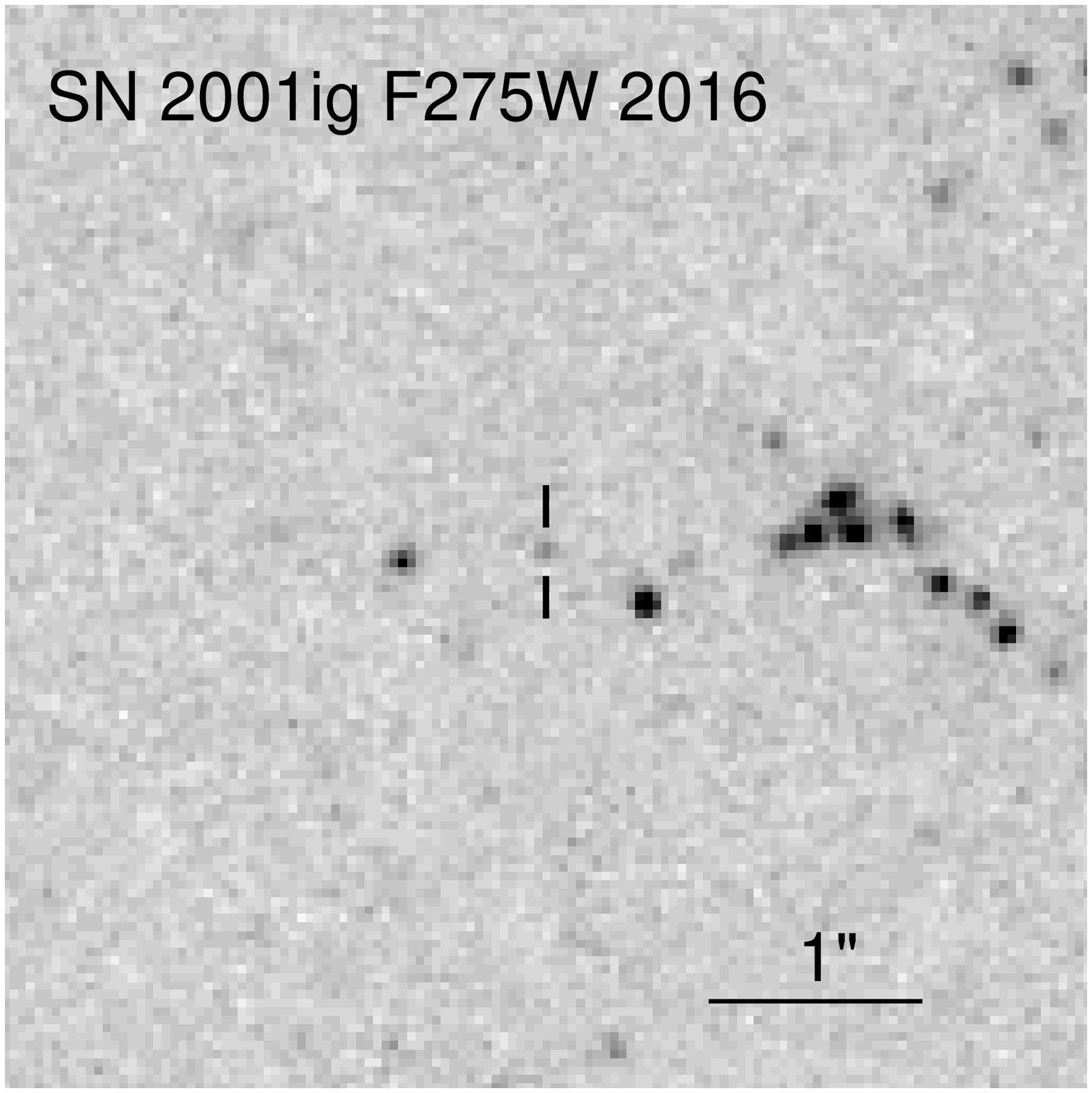}{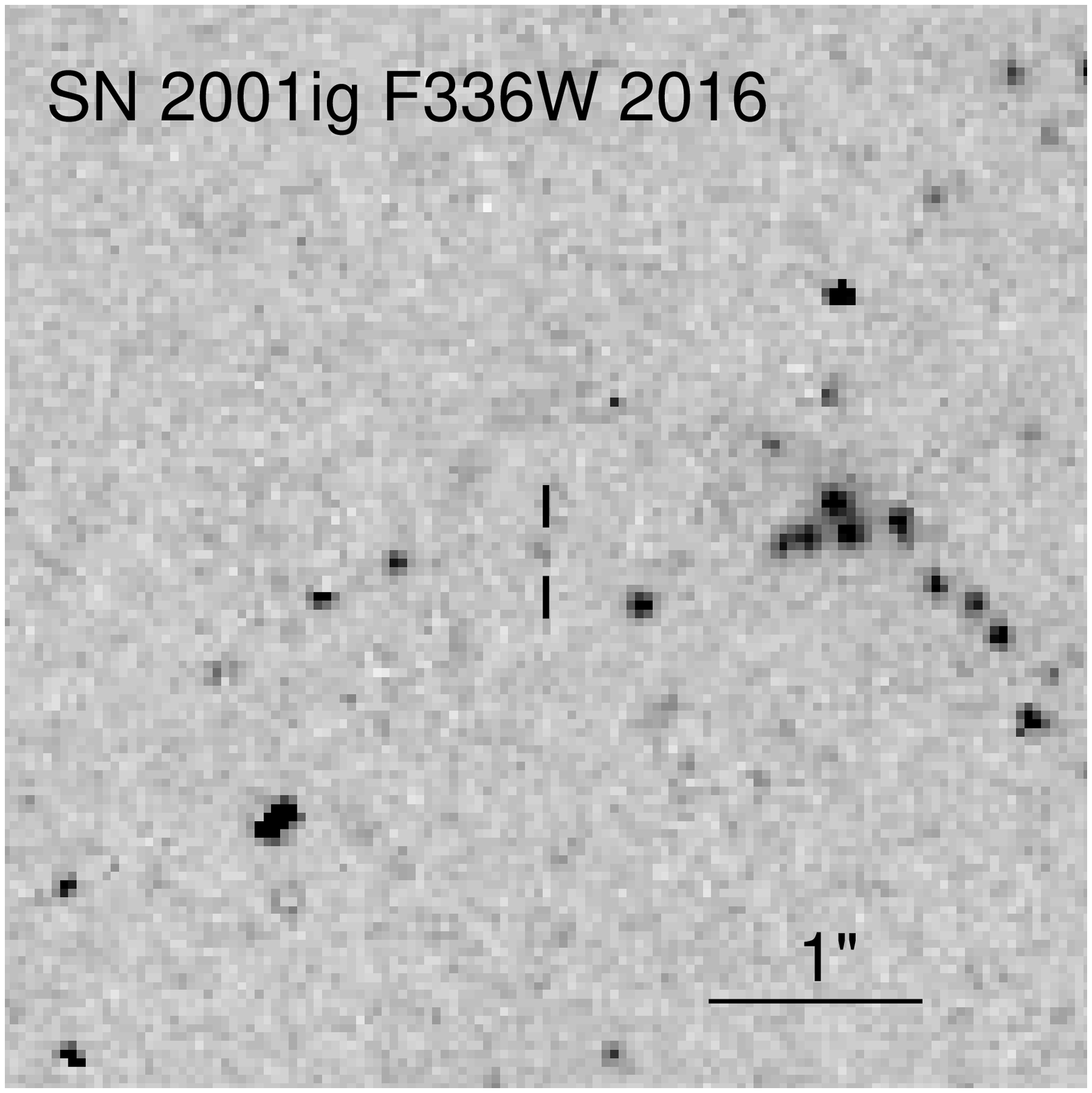}
\caption{{\it (Left)}: A portion of the {\sl HST\/} WFC3/UVIS F275W
image from 2016 obtained as part of GO-14075; the exact site of
SN~2001ig is indicated with tick marks. {\it (Right)}: The same, but 
in F336W. Some cosmic ray hits have not been completely removed from
the image in the right panel. North is up, east to left, in both
panels.\label{figsn}}
\end{figure}

We identified the SN on the WFC3 images by astrometrically aligning 
the F336W image mosaic to the GMOS $g^{\prime}$ image obtained in 
2004 under very good seeing conditions
\citep[$0{\farcs}35$--$0{\farcs}45$ FWHM;][]{Ryder2006}. Using the
eleven stars in common between the two images as astrometric 
fiducials, we identify a source corresponding to the location of 
SN~2001ig with a rms uncertainty of 0.32 UVIS pixel 
(12.7 milliarcsec). The location of the source in both the F275W and 
F336W images is indicated in Figure~\ref{figsn}.  


We measured photometry for this detected source by inputting 
the individual {\em flc} frames into Dolphot \citep{Dolphin00}.  
We ran this routine with parameters set to FitSky=3, RAper=8, and
InterpPSFlib=1, using the TinyTim model point-spread functions (PSFs).
By running the frames first through AstroDrizzle, cosmic ray hits in
the frames had also been flagged, which is important for accurate
aperture correction. As a result, we found for the identified source
$m_{\rm F275W}=25.39 \pm 0.10$ and $m_{\rm F336W}=25.88 \pm 0.13$ mag.
These are robust detections of a point-like source at the position of
SN~2001ig with signal-to-noise ratios of $\sim$12 and 9, respectively.
The object identifier in the Dolphot output was equal to 1 for both
bands, and the sharpness parameter was quite low: $-0.011$ and 
$-0.007$ for F275W and F336W, respectively.

\vskip 10mm
\subsection{GMOS Spectroscopy}
\label{sec:gmos}

Deep optical spectroscopy of the site of SN~2001ig was obtained with
the Gemini South Telescope using GMOS \citep{Hook2004} as part of
program GS-2006B-Q-11 (PI: S.~Ryder). A total of 5~hours on-source
integration was obtained over the course of two nights in 2007, 
July~18 and November~6~UT, in photometric IQ20 (seeing 
$<0.6^{\prime\prime}$) conditions. The B600 grating was used with a
$0\farcs5$ slit, giving a nominal resolution $R\sim1700$.

The GMOS optical spectral data were reduced using the {\sc gmos}
tasks in V1.10 of the {\sc gemini} package within {\sc iraf}. 
A master bias frame constructed by averaging with 3$\sigma$ 
clipping a series of bias frames was subtracted from all raw
images. Images of a Cu-Ar lamp spectrum were used to wavelength
calibrate the science images and straighten (rectify) them along the 
spatial dimension, while images of a quartz-halogen lamp
spectrum helped correct for sensitivity variations within and 
between the original e2v~CCDs. 

\begin{figure}
\includegraphics[angle=-90, width=8.5cm]{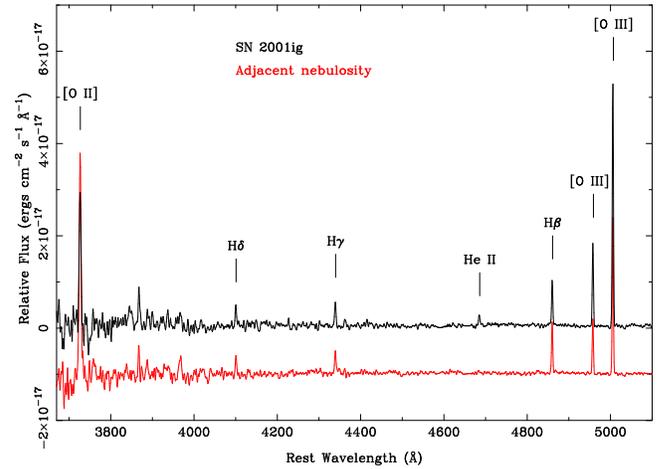}
\caption{Spectra from Gemini/GMOS in 2007 of the site of SN~2001ig
(black), and the nebulosity adjacent to this (red), with the latter
displaced vertically to assist comparison. Major emission features 
are marked. Note the presence of He\,{\sc ii} $\lambda4686$ emission 
only in the spectrum of the SN~2001ig site.\label{snadjspec}}
\end{figure}

The two-dimensional datasets from each night were reduced separately,
then registered spatially and in wavelength space before being
coadded. By comparing continuum sources in the resultant spectral
image with field stars visible in GMOS images from 2004 
\citep[][their Figure 1]{Ryder2006}, the rows containing emission 
from SN~2001ig could
be identified. The flux within an aperture $0\farcs9$ wide centered
on the SN location was extracted, together with an adjacent but not
overlapping aperture of the same width to represent the nebular 
emission from the environment of SN~2001ig. Observations of the
spectrophotometric standard star LTT~9239 \citep{Hamuy1994} with
the same instrument setup and similar airmass to that of the
SN~2001ig observations enabled a system response function to
be derived and applied over the wavelength range 370--650~nm,
as well as the removal of telluric features in the extracted spectra. 
Figure~\ref{snadjspec} compares the spectra at the location of
SN~2001ig and the adjacent nebulosity.

\section{Analysis}

\subsection{Photometric Properties of the Companion}
\label{sec:photres}

The total reddening towards SN 2001ig is uncertain, but we
expect it to be relatively low.  The use of quite a 
narrow slit in order to minimize background contamination of the
SN~2001ig spectrum over several hours of integration (and thus a
large range in slit parallactic angle traversed) precludes the use of
the Balmer decrement to estimate the extinction toward adjacent 
H\,{\sc ii}~regions because of slit losses\footnote{See
\href{http://www.gemini.edu/?q=node/11212}{http://www.gemini.edu/?q=node/11212}.}.
The contribution from the Galactic foreground 
extinction is $A_V=0.029$ mag 
\citep[][via the NASA/IPAC Extragalactic Database, NED]{Schlafly2011}. 
\citet{Silverman2009} measured from their highest resolution spectrum
an equivalent width of $\lesssim 0.1$~\AA\ for the Na~I~D feature and
concluded that the contribution from the host galaxy was likely no
greater than the Galactic contribution; while also noting that
\citet{Maund2007} did require some additional reddening, beyond the
Galactic component, to account for the observed optical polarization
of the SN signal.  We therefore assume that the host reddening is
essentially equivalent to the Galactic reddening; the
total extinction to the SN site is thus $A_V=0.06$ mag.  We further 
assume a \citet{Cardelli1989} reddening law with $R_V=3.1$.

Regarding the metallicity at the site of SN~2001ig, 
\citet{Modjaz2011} found that the oxygen abundance
12+log(O/H) is between 8.27 and 8.53, depending on the abundance diagnostic used. Assuming that solar abundance is 8.7 
\citep[e.g.,][]{Scott2009}, the metallicity appears to be 
around 1/3--2/3 solar at the location of SN~2001ig. 

We analyzed the photometry in the F275W and F336W bands using
\citet{Castelli2003} model atmospheres for main sequence stars at 
the appropriate metallicity ([Fe/H] = $-0.25$) and reddened by the
amount assumed above.  We find that the photometry is consistent 
with an early B-type star with effective temperature $T_{\rm eff}
= $19,000--22,000~K. The corresponding brightness in $V$ of these
models is then 27.11--27.35 mag.

The distance to the host galaxy NGC~7424 is not well determined.
However, it likely sits somewhere between 10.9 \citep{Boker2002} 
and 11.5 Mpc \citep{Tully1988,Soria2006}, depending on the value 
of the Hubble constant assumed, with a corresponding distance 
modulus of 30.19--30.30 mag. If the detected source is a main
sequence star, its absolute magnitude is then in the range
$M_V=-3.25$ to $-2.90$. For the above range in $T_{\rm eff}$, 
at the assumed metallicity, the $V$-band bolometric correction 
is $-1.82$ to $-2.16$ mag
\citep{Paxton2011,Paxton2013,Paxton2015,Choi2016}.
This would imply that the bolometric magnitude is 
$M_{\rm bol}=-5.06 \pm 0.35$ mag,
which for $M_{\rm bol}(\odot)=4.74$ mag, corresponds to a bolometric luminosity 
$\log (L_{\rm bol}/L_{\odot})=3.92 \pm 0.14$.

\begin{figure}
\plotone{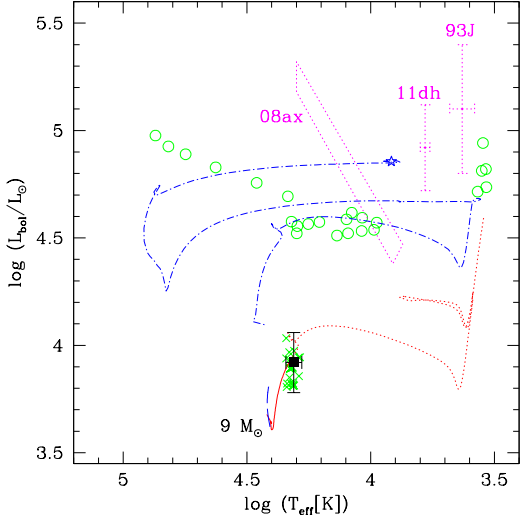}
\caption{Hertzsprung-Russell diagram showing the locus of the point
source detected at the site of SN 2001ig (solid symbol). Its
properties are inferred from comparison of the observed 
{\sl HST\/} ultraviolet photometry with stellar atmosphere models
for main sequence stars at metallicity [Fe/H] = $-0.25$
\citep{Castelli2003}. Shown for comparison in red is a single-star
evolutionary track at 9~$\Msun$ for this same metallicity from MIST.
Also shown are the locations of the secondaries (green crosses)
which are consistent with the detected point source (to within the
uncertainties) of 24 BPASS v2.1 binary evolution models, as well 
as the
corresponding endpoints of the model primaries (green open circles).
For reference, the loci of the progenitors of SN 1993J
\citep{Aldering1994,VanDyk2002}, SN 2011dh
\citep{Maund2011,VanDyk2011}, and SN 2008ax \citep{Folatelli2015}
are shown in magenta (dotted lines). Additionally, for comparison 
in blue is a BPASS binary evolution model with a primary 
(dashed-dotted line) and secondary (long-dashed line) of initial
masses 13 and 9~$\Msun$, respectively, and an initial period of
400~days; the terminus of the primary track is indicated with a
star.\label{fighrd}}
\end{figure}

In Figure~\ref{fighrd} we have placed this inferred $T_{\rm eff}$ 
and $L_{\rm bol}$ on a Hertzsprung-Russell diagram (HRD).
For comparison we show a MIST single-star evolutionary track at
initial mass 9 $\Msun$ at metallicity [Fe/H]  = $-0.25$
\citep{Paxton2011,Paxton2013,Paxton2015,Choi2016};
the track is shown as a solid line up to the data point and 
then extrapolated as a dotted line for the remainder of 
the track. The locus of the detected source agrees with a star at
this mass nearing the terminal-age main sequence (TAMS).

We have found 24 binary evolution models (out of a total of
12678 models generated at this metallicity) from 
BPASS\footnote{\href{http://bpass.auckland.ac.nz/}{http://bpass.auckland.ac.nz/}.}
version 2.1 \citep{Eldridge2017} for which the secondary star 
tracks place them within the uncertainties of the detected object at
the time their primaries explode. In Figure~\ref{fighrd} we 
also show as green circles the endpoints of the tracks of the
corresponding model primaries, i.e. when carbon burning ends,
and core-collapse is imminent. However, in nearly all of these 
cases the primary star endpoint is either cooler (appearing as a 
red supergiant) or hotter than expected for SN~IIb progenitors
including SN~1993J and SN~2011dh (to which SN 2001ig's early 
spectral evolution is most similar:
\citealt{Ryder2006,Shivvers2013}), or even SN~2008ax. This is
evident also in Fig.~21 of \citet{Eldridge2017} which shows
the BPASS models for Type~IIb SNe favoring blue or red
supergiant progenitors over yellow supergiants.

We found one BPASS model having a secondary of initial mass 
9~$\Msun$ (similar to what we infer for the detected source), 
a primary of initial mass 13~$\Msun$ (see Figure~\ref{fighrd}), 
and an initial orbital period of 400~days. Although the agreement 
is not perfect, the endpoint of this model has the secondary on 
the HRD at a locus not far ($\sim 2\sigma$) 
from that of the detected object; while the track of the 13~$\Msun$ 
primary ends at a $T_{\rm eff}$ and $L_{\rm bol}$ that is not too 
dissimilar from the locus of the progenitor of SN 2011dh, which had 
an initial mass 10--19~$\Msun$ \citep{VanDyk2011,Maund2011}); 
or that for SN~1993J \citep{VanDyk2002}, whose progenitor mass was 
in the range 13--22~$\Msun$. 

This particular BPASS model terminates with a primary core mass of 
$\sim$3.5~$\Msun$ and with $\sim$0.04~M$_{\odot}$ of H remaining.
These numbers are quite similar to those yielded by the independent
binary evolution models of \citet{Ouchi2017} for a secondary-to-primary
mass ratio of $0.6 < q < 0.8$, initial orbital period $200 < P < 600$
days, and a low efficiency ($f \approx 0$) of mass accretion from the 
primary onto the secondary. The blue loop in Fig.~\ref{fighrd} is 
similar to that seen in the models of \citet{Yoon2017} for Type~IIb
interacting binary progenitors with periods of a few hundred days, 
and masses closer to 10~M$_{\odot}$ than 20~M$_{\odot}$.
These primaries leave low mass helium-rich remnants (roughly
2--4~M$_{\odot}$) that undergo a second mass transfer stage when 
they swell up again during helium shell burning. Building on the
convention of \citet{Dewi2002} they refer to this as ``Case EBB/LBB''
mass transfer.

The primary in this BPASS model also appears to experience a sudden 
increase in mass loss in the 
last $\sim$50,000~yr, with strongly variable mass loss over the final
$\sim$1700 yr of the model. This appears to be a real effect, not
simply an artifact of the model, and is due to late Roche lobe overflow 
of the He star primary onto the secondary, which changes the surface
conditions and thus the assumed stellar wind mass loss rates (J.~J. 
Eldridge 2017, private communication). Whether such fluctuations in 
mass loss occur in reality is yet to be proven, but would not be
inconsistent with what was inferred for the progenitor from analysis 
of the radio observations of SN~2001ig \citep{Ryder2004}.

Note that we are suggesting here only that this particular model 
has properties that are consistent with that of a SN~IIb progenitor
system, not that this is the actual model for SN~2001ig. We caution
that the endpoint for any given primary star mass depends much more
heavily on the mass loss rates assumed than on the parameters of
the binary system. The relative fraction of primaries that finish 
up as blue supergiants rather than red supergiants increases with
mass loss rate. Mass loss is a notoriously complex problem
\citep[e.g.,][]{Smith2011,Smith2014b,Renzo2017}, and observations
by \citet{Beasor2016} suggest mass loss rates may evolve more 
steeply with luminosity than standard models predict. For
reference the BPASS models adopt mass loss rates for OB stars
from the radiative transfer calculations of \citet{Vink2001},
while for later types they employ the tabulation by 
\citet{deJager1988} of empirical mass loss rates as a function
of position within the HR diagram; in both cases these are 
then scaled appropriately for the metallicity of the star. 
Similarly the Type~IIb binary progenitor models of 
\citet{Yoon2017} use the 
{\tt MESA}\footnote{\href{http://mesa.sourceforge.net}{http://mesa.sourceforge.net}.} code \citep{Paxton2011}, and the
so-called ``Dutch'' scheme for mass loss which is a hybrid of 
these and other mass loss prescriptions \citep{Glebbeek2009}. 
In none of these models however is the mass loss rate tied 
directly to a specific core- or shell-burning phase.


\subsection{Spectroscopic Properties of the Companion}
\label{sec:specres}

Figure~\ref{snadjspec} compares the extracted spectrum from the
location of SN~2001ig with that of the adjacent nebulosity. The 
two are virtually identical, indicating some degree of probable
foreground and/or background contamination of the SN$+$companion
optical spectrum by this nebulosity. The one notable difference
between them, however, is the clear detection (30$\sigma$) of
emission from \heii\ in the SN spectrum only. This narrow line 
(FWHM = 250~km~s$^{-1}$) has a flux ratio relative to the nearby
H$\beta$ line (thus independent of the reddening assumed) of
$0.30\pm0.03$.

Nebular \heii\ emission requires quite hard ionizing radiation, 
such as from an active galactic nucleus, shocks, or X-ray binaries.
It can also be the signpost of the hottest stars, in particular,
Wolf-Rayet (WR) stars, albeit as a much broader feature owing to 
the fast, dense stellar winds \citep{Crowther2007}. Nebular 
\heii\ can also be seen without the other usual accompanying
spectral features of WR stars, \citep[e.g.,][]{LopezSanchez2010}. 
WR stars are claimed to be the progenitors to some types of SESNe
\citep{Smartt2009}, including the Type~IIb SN~2008ax
\citep{Crockett2008}, although \citet{Folatelli2015} use 
post-explosion {\sl HST\/} imaging to argue against a WR progenitor
in that particular case.

Such weak and narrow \heii\ emission was also seen to emerge in 
SN~2014C 1--2 years post-explosion 
\citep{Milisavljevi2015,Anderson2017}.
This is also the interval during which SN~2014C apparently evolved
from an ordinary SN~Ib into a strongly interacting SN~IIn-like
event. In addition, both SN~1993J \citep{Matheson2000} and 
SN~2013df \citep{Maeda2015} underwent a spectral transformation
after the first year, indicative of interaction with a dense CSM,
albeit through the appearance of broad, flat-topped H$\alpha$ and
He\,{\sc i} lines, rather than \heii. Transitory \heii\ emission 
was also seen briefly at around 100~days in the Type Ibn SN~2006jc
\citep{Smith2008}, coincident with the onset of dust formation in 
a dense CSM shell.

Optical spectroscopy by \citet{Milisavljevi2015}, as well as 
radio observations by \citet{Anderson2017} and X-ray observations by
\citet{Margutti2017}, led to the conclusion that the progenitor of
SN~2014C initially exploded into a low-density cavity
before eventually running into a dense, H-rich shell of material
shed long before it exploded. Such delayed interaction has been
seen previously in objects such as SN~1996cr \citep{Bauer2008}, as
well as SN~1987A, which has slowly overtaken its dense, clumpy ring.
Thus, it may be that this \heii\ emission in SN~2001ig is a signpost
of ongoing or renewed interaction with CSM much farther out than
that probed by the radio emission. This would suggest
that the SN~2001ig progenitor, like the case of SN~2014C, may have
been intermediate between the more compact progenitor (with less
dense CSM) of SN~2011dh, and the more extended progenitor (and
stronger CSM interaction) of SN~1993J.

\section{Discussion}
\label{sec:disc}

\subsection{An Innocent Bystander?}
\label{sec:bystander}

While the object detected in the WFC3 images could simply be a
chance alignment of an unrelated foreground/background source, this
seems extremely unlikely for two reasons. First, the somewhat sparse
distribution of detectable UV sources in Figure~\ref{figsn} makes this
a rather low statistical probability; and second, it is as unlikely
that this contaminating source just happens to show rare, narrow \heii\
emission, rather than a much more common stellar continuum.

On the basis of ground-based optical imaging obtained in 2004,
\citet{Ryder2006} suggested that this surviving companion was most likely
a supergiant star, somewhere between a late-B (for $A_V \sim 1$~mag) to
late-F (for $A_V=0$ mag) spectral type. Our {\sl HST\/} UV imaging taken 
12~years later now indicates the companion is more likely to be a main
sequence B star with quite low extinction ($A_V=0.06$ mag;
Section~\ref{sec:photres}). We found that the ($g^{\prime} - r^{\prime}$)
color and ($u^{\prime} - g^{\prime}$) blue limit of the 2004 source were
inconsistent with either a pure H\,{\sc ii}~region or the nebular spectrum
of SN~1993J at a similar age. It is quite likely that the source spectrum
will have evolved since 2004, fading in the manner extrapolated for
SN~2011dh by \citet{Maund2015}, but perhaps rebrightening owing to 
ongoing or resumed CSM interaction as hinted at by the emergence of 
\heii\ emission. Unfortunately, there has been very little photometric 
monitoring of SN~2001ig throughout its evolution to
yield a meaningful light curve from which to extrapolate to such late
times.

\subsection{Binary Evolutionary Channel for the SN 2001ig Progenitor}
\label{sec:binchan}

The presence of a surviving stellar companion would imply that 
the progenitor star of SN~2001ig was in a binary system. Indeed, 
if that is the case, mass transfer onto the binary companion
of the progenitor would have played an important role in the 
removal of almost all of its H-rich envelope, thus leading
to a Type~IIb SN event. 

\citet{Zapartas2017a} provide theoretical predictions for the
binary companions of all SESNe (including SNe~IIb
and SNe~Ib/c). They find that in the majority of scenarios a binary
companion star is expected at the moment of explosion. In most cases
the companion still resides on the main sequence, since it
evolves on a longer timescale owing to its lower initial mass.
In fact, our inferred mass of $\sim$9~$\Msun$
for the companion of SN 2001ig coincides with the broad
peak of the predicted mass distribution of main sequence companions of all
SESNe shown in \citet{Zapartas2017a}.

Binary channels that originate from short-period systems
(about $1 \lesssim \log P ({\rm days}) \lesssim 3$) have been
suggested to result in compact SN~IIb progenitors
\citep{Yoon2010,Stancliffe2009,Bersten2012,Benvenuto2013,Folatelli2015,
Yoon2017}, with a thin H envelope of mass 
$\lesssim 0.15 \Msun$ \citep[e.g.,][]{Yoon2017}. 
In low-metallicity environments, weak stellar winds allow the
thin remaining envelope to still be present when the progenitor
explodes. The progenitors of these compact SNe~IIb stay on the 
blue part of the HRD or are expected to end their lives as yellow
supergiants (YSGs), as in the case of direct progenitor detections
of SN~2011dh and SN~2013df. The BPASS model shown in
Figure~\ref{fighrd} follows such a scenario, having an initial
period $\log P = 2.6$ and leaving 0.04~$\Msun$ of H
remaining on the progenitor at the moment of explosion.

Alternatively, wider initial orbits lead to more massive H
envelopes \citep[e.g.,][]{Yoon2017}. Progenitors with an extended
low-mass H envelope of $\sim$0.1--0.5~$\Msun$
\citep{Podsiadlowski1993,Woosley1994, Elmhamdi2006,Claeys2011} 
are expected to stay on the red part of the HRD.
\citet{Claeys2011} find that extended SN~IIb progenitors
originate from almost equal-mass systems with 
$q=M_{\rm accretor}/M_{\rm donor} \gtrsim$~0.7--0.8 in initially
wide orbits of $ 3 \lesssim \log P ({\rm days}) \lesssim 3.3$.
Although they assume conservative mass transfer as their standard
assumption, they also explored binary evolution for different 
mass-transfer efficiencies to determine the uncertainty in how much mass 
is accreted by the companion star. 

In the case of low efficiency, as expected in systems with wide orbits 
\citep[e.g.,][]{Schneider2015}, the inferred companion mass 
of $\sim$9~$\Msun$ is relatively close to its birth mass. At the
same time, the progenitor star initially should be somewhat more massive,
since $q \gtrsim 0.7$, and will naturally form a He core of
3--4~$\Msun$, consistent with the ejecta mass of $\sim$1.15~$\Msun$
inferred by \citet{Silverman2009}. Thus, in the scenario
of an extended SN~IIb discussed here, the small difference in the
evolutionary timescales owing to the similarity of the mass, as 
well as the possible absence of significant mass accretion onto
the companion, results in only a limited rejuvenation 
of the companion \citep{Hellings1983,Hellings1984}.
This could explain the fact that the detected companion appears
to lie close to the TAMS, as seen in Figure~\ref{fighrd}.

A further consequence of possible non-conservative mass transfer
is that SN~IIb progenitors are expected to have significant
CSM around them just before the explosion, corresponding to 
mass loss rates on the order of $10^{-5}$ to 
$10^{-4} \Msun~{\rm yr}^{-1}$, when averaged over the final 
1000 years before the SN explosion \citep{Ouchi2017}, which is
comparable to that inferred from the radio observations for 
SN 2001ig \citep{Ryder2004}. The CSM produced by the binary
interaction could be the cause of the observed \heii\ feature
discussed in Section~\ref{sec:specres}.
%

In summary, constraining the exact parameters of the original
progenitor binary system is difficult owing to the lack of
extensive optical photometric monitoring of SN~2001ig, or a direct
detection of its progenitor. If SN~2001ig is similar to SN~2011dh, 
the expected YSG progenitor could have
formed through mass transfer in a binary system with an initial
period of a few hundred days, as is the case for the BPASS model shown in
Figure~\ref{fighrd}. However, we cannot exclude the possibility
of wider initial orbits which would lead to more extended SN~IIb
progenitors.

\section{Summary and Conclusions}
\label{sec:con} 

We have obtained late-time UV imaging (over 14~years past explosion) 
of the site of SN~2001ig and identified in two separate filters a point
source at the known location of the SN. Allowing for the
uncertainties in distance and extinction toward the host galaxy, 
we find this source to be consistent with an early B-type main
sequence star with $T_{\rm eff} = $19,000--22,000~K and a bolometric
luminosity $\log (L_{\rm bol}/L_{\odot})=3.92 \pm 0.14$. We show that
the evolutionary track of a BPASS model with a 9~\Msun\ secondary star
passes near the TAMS on the HRD at about the
same time that its 13~\Msun\ primary companion reaches a location on the
HRD similar to those observed for the progenitors of SN~2011dh
and SN~1993J. The growing number of surviving companions found
in SNe~IIb, coupled with their relatively low progenitor
masses, weakens the case for massive single stars such as 
LBVs being the progenitors of most SNe~IIb
\citep{Soderberg2006,Kotak2006,Groh2013}.

Although the progenitor star of SN~2001ig was never 
identified directly, we believe our detection of what is almost
certainly the surviving companion in this model makes a strong case 
for a binary interaction scenario for SN~2001ig, leading to a 
partially-stripped envelope and dense CSM.
A ground-based optical spectroscopic comparison of the location of
SN~2001ig with its neighborhood at an age of almost 6~years reveals
narrow \heii\ emission, which we interpret as further evidence for
interaction with this dense CSM.

SN~2001ig initially underwent strong interaction of the SN shock with
the pre-existing CSM surrounding the progenitor for years after
explosion, as revealed by its long-lived radio emission. This is a
common property of SNe~IIb, such as SN 1993J \citep{Weiler2007}, 
SN 2011dh \citep{Horesh2013}, and SN 2013df \citep{Kamble2016}.
In the case of SN 2011dh, \citet{Maund2015} discussed the possibility 
of ongoing or renewed CSM interaction accounting for its observed 
late-time, {\sl HST}-based optical spectral energy distribution. 
These authors pointed out that
the UV emission from SN~2011dh, interpreted by \citet{Folatelli2014}
as evidence for a binary companion, could instead be associated with 
such a potentially protracted CSM interaction, and recommended further 
monitoring at optical wavelengths with {\sl HST\/} once this still
relatively young SESN has faded enough to distinguish between these 
two possibilities. Similarly, additional optical imaging with 
{\sl HST\/} in the future will be necessary to determine whether the 
UV emission we have detected at the site of SN~2001ig is produced
entirely by a surviving hot companion, or has a contribution from 
long-term, low-level CSM interaction.

\section*{Acknowledgments}
%
This work is based in part on observations made with the NASA/ESA 
{\it Hubble Space Telescope}, obtained at the Space Telescope Science
Institute (STScI), which is operated by the Association of Universities
for Research in Astronomy, Inc., under NASA contract NAS 5-26555.
Support was provided by NASA through grants GO-14075 and AR-14295
from STScI. 
It is also based in part on observations obtained at the Gemini
Observatory, which 
is operated by the Association of Universities for Research in
Astronomy, Inc., under a cooperative agreement with the NSF on 
behalf of the Gemini partnership: the National Science Foundation
(United States), the National Research Council (Canada), CONICYT
(Chile), Ministerio de Ciencia, Tecnolog\'{i}a e Innovaci\'{o}n
Productiva (Argentina), and Minist\'{e}rio da Ci\^{e}ncia, 
Tecnologia e Inova\c{c}\~{a}o (Brazil).
We thank the referee for their suggestions, and are grateful to
J.~J.~Eldridge for discussions regarding the BPASS models. AVF's group
is also grateful for generous financial assistance from the Christopher
R. Redlich Fund, the TABASGO Foundation, NSF grant AST-1211916, and the
Miller Institute for Basic Research in Science (U.C. Berkeley).
EZ is supported by a grant of the Netherlands Research School for
Astronomy (NOVA).  SdM acknowledges support by a Marie 
Sklodowska-Curie Action (H2020 MSCA-IF-2014, project
BinCosmos, id 661502).  


\bibliographystyle{apj}
\bibliography{sn2001ig}

\end{document}